\documentclass[superscriptaddress,aps,prl,showkeys,showpacs,twocolumn,10pt]{revtex4-1} 
\usepackage{graphicx}
\usepackage{amsmath}
\usepackage{amssymb}
\usepackage{bm}
\usepackage[OT1,T1]{fontenc}
\usepackage{dcolumn}

\newcolumntype{d}[1]{D{.}{.}{#1}}

\begin{document}
\title{Investigation of the $dd \to \mathrm{^3He}n\pi^{0}$ reaction with the FZ J\"{u}lich WASA-at-COSY facility}

\date{July 22, 2013}

\newcommand*{\IKPUU}{Division of Nuclear Physics, Department of Physics and 
 Astronomy, Uppsala University, Box 516, 75120 Uppsala, Sweden}
\newcommand*{\ASWarsN}{Department of Nuclear Physics, National Centre for 
 Nuclear Research, ul.\ Hoza~69, 00-681, Warsaw, Poland}
\newcommand*{\IPJ}{Institute of Physics, Jagiellonian University, ul.\ 
 Reymonta~4, 30-059 Krak\'{o}w, Poland}
\newcommand*{\PITue}{Physikalisches Institut, Eberhard--Karls--Universit\"at 
 T\"ubingen, Auf der Morgenstelle~14, 72076 T\"ubingen, Germany}
\newcommand*{\Kepler}{Kepler Center for Astro and Particle Physics, Eberhard 
 Karls University T\"ubingen, Auf der Morgenstelle~14, 72076 T\"ubingen, 
 Germany}
\newcommand*{\MS}{Institut f\"ur Kernphysik, Westf\"alische 
 Wilhelms--Universit\"at M\"unster, Wilhelm--Klemm--Str.~9, 48149 M\"unster, 
 Germany}
\newcommand*{\ASWarsH}{High Energy Physics Department, National Centre for 
 Nuclear Research, ul.\ Hoza~69, 00-681, Warsaw, Poland}
\newcommand*{\IITB}{Department of Physics, Indian Institute of Technology 
 Bombay, Powai, Mumbai--400076, Maharashtra, India}
\newcommand*{\IKPJ}{Institut f\"ur Kernphysik, Forschungszentrum J\"ulich, 
 52425 J\"ulich, Germany}
\newcommand*{\JCHP}{J\"ulich Center for Hadron Physics, Forschungszentrum 
 J\"ulich, 52425 J\"ulich, Germany}
\newcommand*{\Bochum}{Institut f\"ur Experimentalphysik I, Ruhr--Universit\"at 
 Bochum, Universit\"atsstr.~150, 44780 Bochum, Germany}
\newcommand*{\ZELJ}{Zentralinstitut f\"ur Engineering, Elektronik und 
 Analytik, Forschungszentrum J\"ulich, 52425 J\"ulich, Germany}
\newcommand*{\Erl}{Physikalisches Institut, 
 Friedrich--Alexander--Universit\"at Erlangen--N\"urnberg, 
 Erwin--Rommel-Str.~1, 91058 Erlangen, Germany}
\newcommand*{\ITEP}{Institute for Theoretical and Experimental Physics, State 
 Scientific Center of the Russian Federation, Bolshaya Cheremushkinskaya~25, 
 117218 Moscow, Russia}
\newcommand*{\Giess}{II.\ Physikalisches Institut, 
 Justus--Liebig--Universit\"at Gie{\ss}en, Heinrich--Buff--Ring~16, 
 35392 Giessen, Germany}
\newcommand*{\IITI}{Department of Physics, Indian Institute of Technology 
 Indore, Khandwa Road, Indore--452017, Madhya Pradesh, India}
\newcommand*{\HepGat}{High Energy Physics Division, Petersburg Nuclear Physics 
 Institute, Orlova Rosha~2, Gatchina, Leningrad district 188300, Russia}
\newcommand*{\IAS}{Institute for Advanced Simulation, Forschungszentrum 
 J\"ulich, 52425 J\"ulich, Germany}
\newcommand*{\HISKP}{Helmholtz--Institut f\"ur Strahlen-- und Kernphysik, 
 Rheinische Friedrich--Wilhelms--Universit\"at Bonn, Nu{\ss}allee~14--16, 
 53115 Bonn, Germany}
\newcommand*{\HiJINR}{Veksler and Baldin Laboratory of High Energiy Physics, 
 Joint Institute for Nuclear Physics, Joliot--Curie~6, 141980 Dubna, Moscow 
 region, Russia}
\newcommand*{\Katow}{August Che{\l}kowski Institute of Physics, University of 
 Silesia, Uniwersytecka~4, 40-007, Katowice, Poland}
\newcommand*{\IFJ}{The Henryk Niewodnicza{\'n}ski Institute of Nuclear 
 Physics, Polish Academy of Sciences, 152~Radzikowskiego St, 31-342 
 Krak\'{o}w, Poland}
\newcommand*{\NuJINR}{Dzhelepov Laboratory of Nuclear Problems, Joint 
 Institute for Nuclear Physics, Joliot--Curie~6, 141980 Dubna, Moscow region, 
 Russia}
\newcommand*{\KEK}{High Energy Accelerator Research Organisation KEK, Tsukuba, 
 Ibaraki 305--0801, Japan}
\newcommand*{\IMPCAS}{Institute of Modern Physics, Chinese Academy of 
 Sciences, 509 Nanchang Rd., Lanzhou 730000, China}
\newcommand*{\ASLodz}{Department of Cosmic Ray Physics, National Centre for 
 Nuclear Research, ul.\ Uniwersytecka~5, 90--950 {\L}\'{o}d\'{z}, Poland}

\newcommand*{\Delhi}{Department of Physics and Astrophysics, University of 
 Delhi, Delhi--110007, India}
\newcommand*{\SU}{Department of Physics, Stockholm University, 
 Roslagstullsbacken~21, AlbaNova, 10691 Stockholm, Sweden}
\newcommand*{\Mainz}{Institut f\"ur Kernphysik, Johannes 
 Gutenberg--Universit\"at Mainz, Johann--Joachim--Becher Weg~45, 55128 Mainz, 
 Germany}
\newcommand*{\UCLA}{Department of Physics and Astronomy, University of 
 California, Los Angeles, California--90045, U.S.A.}
\newcommand*{\Bern}{Albert Einstein Center for Fundamental Physics, 
 Fachbereich Physik und Astronomie, Universit\"at Bern, Sidlerstr.~5, 
 3012 Bern, Switzerland}

\author{P.~Adlarson}    \affiliation{\IKPUU}
\author{W.~Augustyniak} \affiliation{\ASWarsN}
\author{W.~Bardan}      \affiliation{\IPJ}
\author{M.~Bashkanov}   \affiliation{\PITue}\affiliation{\Kepler}
\author{F.S.~Bergmann}  \affiliation{\MS}
\author{M.~Ber{\l}owski}\affiliation{\ASWarsH}
\author{H.~Bhatt}       \affiliation{\IITB}
\author{M.~B\"uscher}   \affiliation{\IKPJ}\affiliation{\JCHP}
\author{H.~Cal\'{e}n}   \affiliation{\IKPUU}
\author{I.~Ciepa{\l}}   \affiliation{\IPJ}
% ...
\collaboration{WASA-at-COSY Collaboration}\noaffiliation

\begin{abstract}
An exclusive measurement of the $dd \to \mathrm{^3He}n\pi^{0}$ reaction was 
carried out at a beam momentum of $p_{d}=$~1.2 GeV/c using the WASA-at-COSY 
facility. Information on the total cross section as well as 
differential distributions was obtained. The data are described by a 
phenomenological approach based on a combination of a quasi-free model 
and a partial wave expansion for the three-body reaction. The total cross 
section is found to be 
$\sigma_{tot}= (2.89 \pm 0.01_{stat}\pm 0.06_{sys} \pm 0.29_{norm})\,\mathrm{\mu b}$.
The contribution of the quasi-free processes (with the beam or target neutron being 
a spectator) accounts for 38$\%$ of the total cross section 
and dominates the differential distributions in specific regions of
phase space. The remaining part of the cross section can be described 
by a partial wave decomposition indicating the significance of 
$p$-wave contributions in the final state.
\end{abstract}
\maketitle

\section{Introduction}
At the fundamental level of the Standard Model, isospin violation
is due to quark mass differences as well as electromagnetic
effects~\cite{Wein,Gasser,Leut}. Therefore, the observation of isospin 
violation is an experimental tool to study quark mass effects in hadronic 
processes. However, in general isospin violating observables are largely 
dominated by the pion mass differences, which are enhanced due to the small
pion mass. An exception are charge symmetry breaking (CSB) observables. Charge
symmetry is the invariance of a system under rotation by 180$^\circ$
around the second axis in isospin space that interchanges up and down quarks.
It transforms a $\pi^+$ into a $\pi^-$ and, therefore,
the pion mass difference does not contribute. Ref.~\cite{Miller90} calls the 
investigation of CSB effects one of the most challenging subjects in hadron 
physics. On the basis of theoretical approaches with a direct connection to QCD, 
like lattice QCD or effective field theory, it is possible to study
quark mass effects on the hadronic level, since the effects
of virtual photons are under control --- for a detailed discussion
on this subject see Ref.~\cite{Fettes:2000vm}.

The first observation of the charge symmetry breaking $dd \to \mathrm{^4He}\pi^{0}$ reaction 
was reported for 
beam energies very close to the reaction threshold \cite{Stephenson03}. At 
the same time information on CSB in $np \to d \pi^{0}$ manifesting itself in a 
forward-backward asymmetry became available \cite{Opper03}. These data triggered 
advanced theoretical calculations within effective field theory, providing the 
opportunity to investigate the influence of the quark masses in nuclear physics
\cite{miller06,filin}. This is done using Chiral Perturbation Theory (ChPT) 
which has been extended to pion production reactions \cite{Hanhart04}. First 
steps towards a theoretical understanding of the $dd \to \mathrm{^4He}\pi^{0}$ reaction
have been taken \cite{Gardestig04,Nogga06}. Soft photon exchange in the initial 
state could significantly enhance the cross section for 
$dd \to \mathrm{^4He}\pi^{0}$~\cite{timo}. However, it was demonstrated in 
Ref.~\cite{jerrynew} that a simultaneous analysis of CSB in the two-nucleon 
sector and in $dd \to \mathrm{^4He}\pi^{0}$ strongly constrains the calculations.

The main problem in the calculation of $dd\to \mathrm{^4He}\pi^0$ is to get
theoretical control over the isospin symmetric part of the initial state interactions, 
for here high accuracy wave functions are needed for $dd\to 4N$ in low partial waves
at relatively high energies. These can be accessed by measurements of
other, isospin conserving, $dd$ induced pion production reaction
channels at a similar excess energy, such that the final state (and, thus, also
the initial state) is constrained
to small angular momenta. Then, the incoming system shares some of the partial waves 
in the initial state with the reaction $dd\to \mathrm{^4He}\pi^0$, while the transition 
operator is calculable with sufficient accuracy using ChPT.
Such a reaction is $dd\to \mathrm{^3He}n\pi^0$ and the corresponding measurement
is presented here.

\section{Experiment}
The experiment was carried out at the Institute for Nuclear Physics of Forschungszentrum
J\"ulich in Germany using the Cooler Synchrotron COSY \cite{Maier97} together with the 
WASA detection system. For the measurement of $dd \to \mathrm{^3He}n\pi^{0}$ at an excess 
energy of $Q\approx\,$40~MeV a deuteron beam with a momentum of 1.2~GeV/c was scattered on 
frozen deuterium pellets provided by an internal pellet target. The reaction products 
$\mathrm{^3He}$ and $\pi^{0}$ were detected by the Forward Detector and the Central Detector 
of the WASA facility, respectively, while the neutron remained undetected.
The Forward Detector consists of several layers of plastic scintillators
for particle identification and energy reconstruction and an array of straw tubes
for precise tracking. The polar angular range between 3$^\circ$ and 18$^\circ$ fully covers 
the angular range of the outgoing $\mathrm{^3He}$ with the exception
of very small angles. At this beam momentum the $\mathrm{^3He}$ ejectiles have
kinetic energies in the range of 65 - 214~MeV and, thus, 
are already stopped in the first detector layers:
in addition to the straw tube tracker only the two 3~mm thick layers of the Forward Window Counter 
and the first 5 mm thick layer of the Forward Trigger Hodoscope were used. 
The two photons from the $\pi^{0}$ decay were detected by the Scintillator Electromagnetic Calorimeter
as part of the Central Detector. Photons were distinguished from charged particles using 
the Plastic Scintillator Barrel located inside the calorimeter. 
The experiment trigger was based on a coincidence between a high energy deposit in both layers of 
the Forward Window Counter together with a veto condition on the first layer of the Forward Range
Hodoscope to select helium ejectiles and a low energy neutral cluster ($E$ > 20~MeV) in the calorimeter
to tag the decay of the pion. 
Further information on the WASA-at-COSY facility can be found in Ref.\cite{wasa}.

\section{Data Analysis}
\begin{figure}
\begin{center}
\includegraphics[width=\columnwidth]{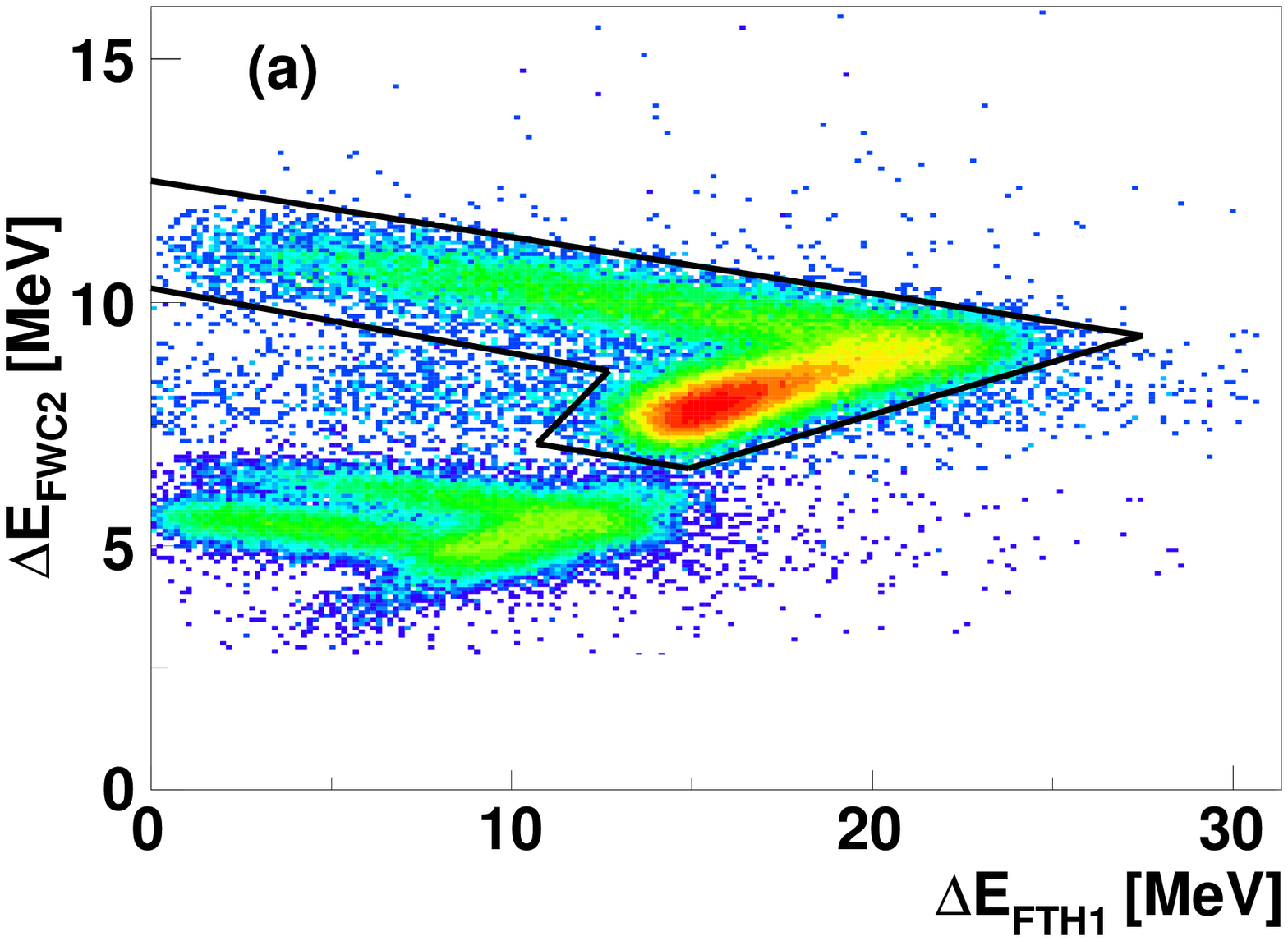} 
\includegraphics[width=\columnwidth]{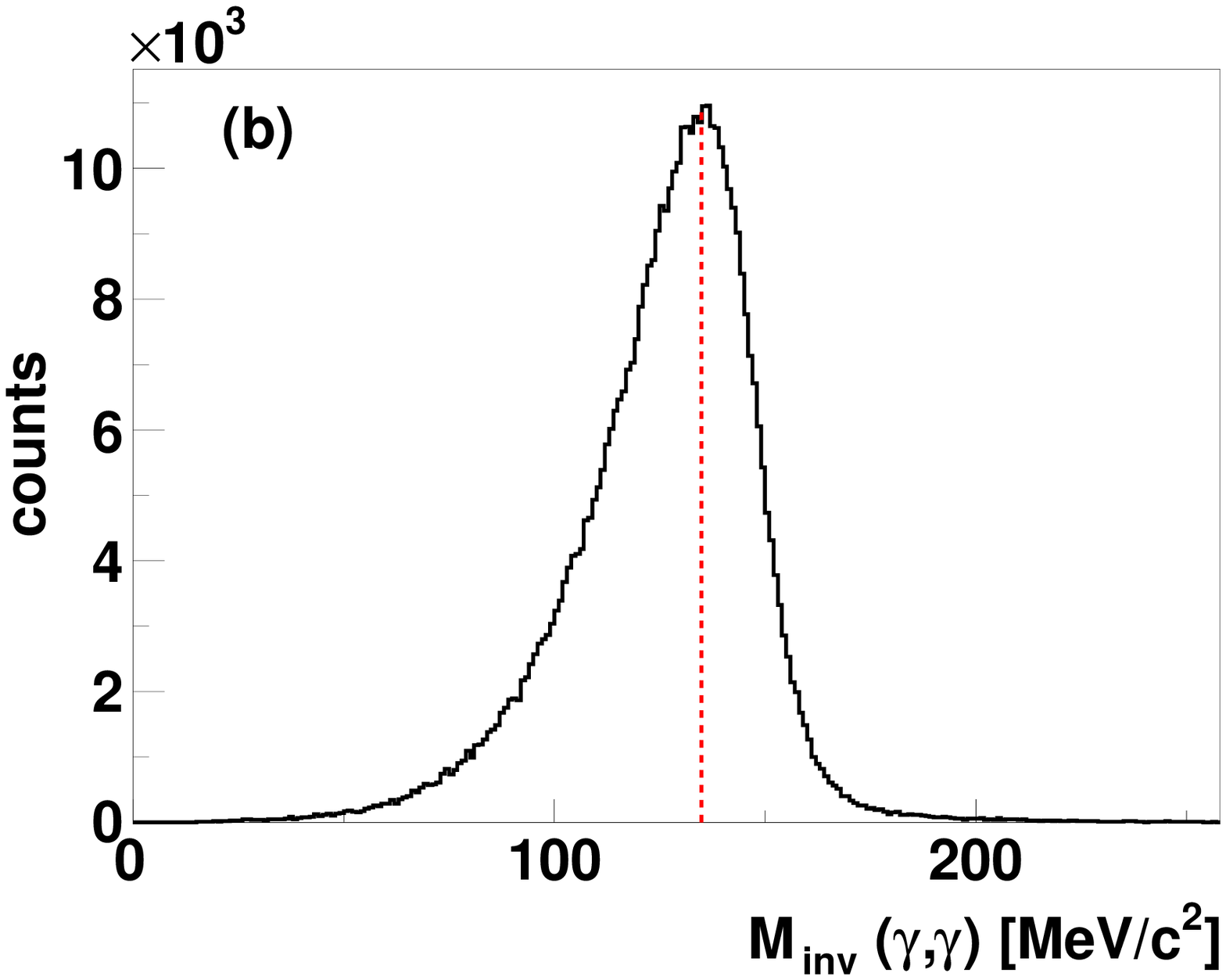}
\end{center}
\caption{(Color online) (a) Energy loss in the Forward Window Counter versus energy loss in the
first layer of the Forward Trigger Hodoscope. The obtained energy pattern shows a clear 
separation between different particles types. The graphical cut indicated in black 
represents the region used to select $\mathrm{^3He}$ candidates. (b) The two photon 
invariant mass distribution corresponding to the $\pi^{0} \to \gamma\gamma$ decay. 
The red dotted line indicates the $\pi^{0}$ mass.
}
\label{fig:de_inv}       
\end{figure}

Apart from the charge symmetry breaking reaction $dd \to \mathrm{^4He}\pi^{0}$ with 
a four orders of magnitude smaller cross section, $dd \to \mathrm{^3He}n\pi^{0}$ is 
the only process with a charge 2 particle and a neutral pion in final state. Thus, 
the identification of a forward going helium and two neutral tracks forming a pion already
provides a clean signature for this reaction. Helium isotopes are identified by 
means of $\Delta E-\Delta E$ plots using the energy deposit in the Forward Window Counter 
and the first layer of the Forward Trigger Hodoscope (Fig.~\ref{fig:de_inv}a). 

The condition that the $\mathrm{^3He}$ has to pass at least the first two scintillator
layers introduces an additional acceptance cut of $E_{kin}$ > 125~MeV. This rejects 
most of the $\mathrm{^3He}$ going backward in the c.m. system. However, having two 
identical particles in the initial state and, thus, a symmetric angular distribution 
with respect to $\theta_\mathrm{c.m.}$ = 90$^\circ$ the full angular range can be recovered by 
a symmetrization of the detected events. 
The energy deposits are also used to reconstruct the $\mathrm{^3He}$ kinetic
energy by matching the energy loss pattern to Monte-Carlo simulations.
The $\mathrm{^3He}$ four-momentum is completed by the direction information from the
straw tube tracker.
In addition to the $\mathrm{^3He}$ two neutral clusters in the central detector corresponding 
to the two photons from the $\pi^{0}$ decay were requested. As event pile-up and
low energy satellites of genuine photon clusters can cause larger 
photon multiplicities the most probable true two-photon combination was
identified by selecting the pair with the $\mathrm{^3He}-\pi^{0}$ missing mass being closest
to the neutron mass. As result a nearly background free pion peak was obtained 
(Fig.~\ref{fig:de_inv}b). In a final step the data were refined by applying a kinematic 
fit using the hypothesis $dd \to \mathrm{^3He}n\pi^{0}$. Still remaining background and 
badly reconstructed events were rejected by a cut on the cumulated probability distribution
at 10$\%$. At the end of the analysis chain about $3.4\times10^{6}$ fully reconstructed 
and background free $dd \to \mathrm{^3He}n\pi^{0}$ events are available.  
Although based on this data set any possible differential distribution can be generated 
--- e.g. for a selective comparison with future microscopic 
theoretical calculations --- a suitable set of observables for further 
analysis and presentation had to be selected. For any unpolarized
measurement with three particles in final state four independent
variables fully describe the reaction kinematics. For the present 
analysis the choice for these independent variables is based
on the Jacobi momenta $\vec{q}$ and $\vec{p}$ with $\vec{q}$ being the $\pi^{0}$ momentum in the overall 
c.m. system and $\vec{p}$ the momentum in the rest frame of the $\mathrm{^3He}-n$ subsystem. The
following variables were constructed accordingly: $\cos\theta_{q}$, $\cos\theta_{p}$
(the polar angles of $\vec{q}$ and $\vec{p}$, respectively),  $M_{\mathrm{^3He}n}$ and $\varphi$ 
(the angle between the projections of $\vec{q}$ and $\vec{p}$ onto the xy-plane). As discussed earlier
all plots show data after a symmetrization in the global c.m. system.

\begin{figure}[b]
\begin{center}
\includegraphics[width=\columnwidth]{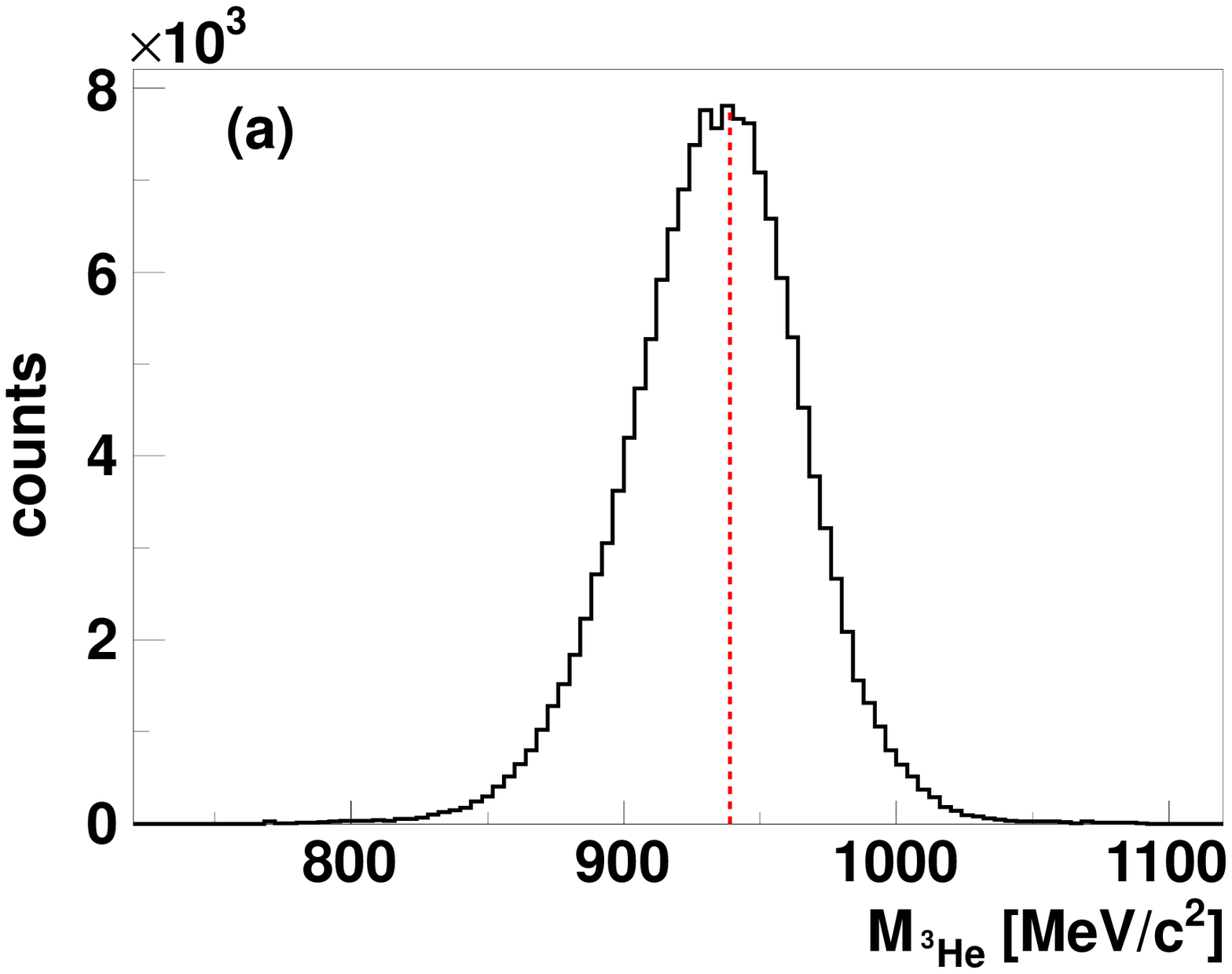} 
\includegraphics[width=\columnwidth]{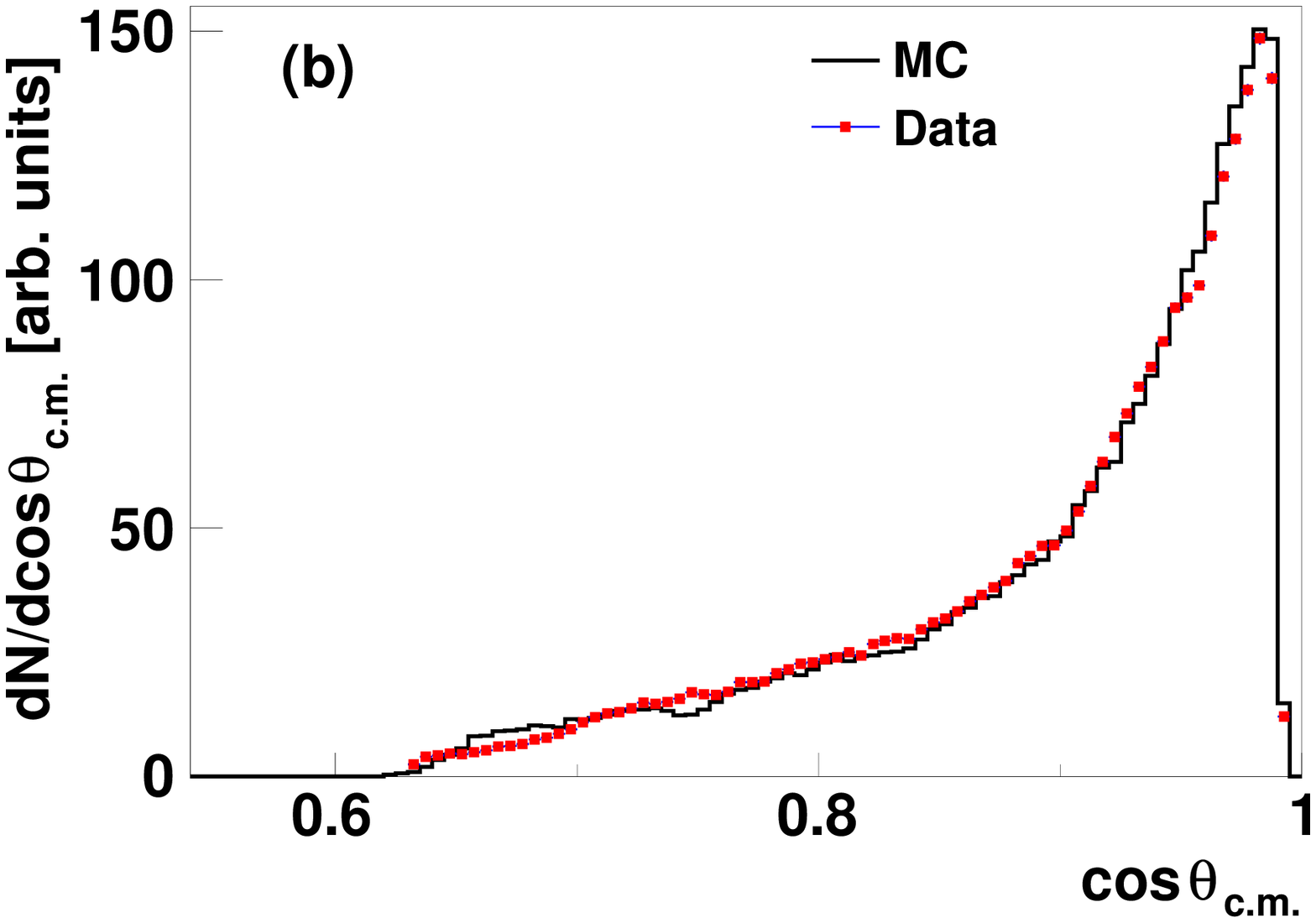} 
\end{center}
\caption{(Color online) Measurement of the $dd \to \mathrm{^3He}n$ reaction. (a) $\mathrm{^3He}$ missing 
mass distribution, the vertical red dotted line indicates the neutron mass. (b) Measured 
angular distribution in comparison with a Monte Carlo simulation based on a parametrized 
cross section (see text).
}
\label{fig:norm}       
\end{figure}

The absolute normalization was done relative to the $dd \to \mathrm{^3He}n$ reaction.
Corresponding data were taken in parallel during the first part of the run using 
a dedicated trigger. Due to the correlation between kinetic
energy and scattering angle for the binary reaction, 
quasi mono-energetic particles form a distinct and clean peak in 
the $\Delta E -\Delta E$ plots. For the selected events the $\mathrm{^3He}$ missing
mass distribution reveals a background free peak at
the mass of the neutron (Fig.~\ref{fig:norm}a). In order to determine the integrated 
luminosity the data presented in Ref.~\cite{Bizard80} were used. 
The authors measured the reaction $dd \to \mathrm{^3H}p$ for beam momenta 
between 1.09~GeV/c - 1.78~GeV/c and $dd \to \mathrm{^3He}n$ for beam momenta 
in the range of 1.65~GeV/c - 2.5~GeV/c. Moreover, they showed  that the differential 
cross sections for both channels at 1.65~GeV/c are identical within the presented errors. 
Based on these results we used the measured cross sections for $dd \to \mathrm{^3H}p$ to 
calculate the cross sections for $dd \to \mathrm{^3He}n$ at 1.2~GeV/c. For this the angular 
distributions for the beam momenta of 1.109~GeV/c, 1.387~GeV/c and 1.493~GeV/c were 
parametrized. Then, for selected polar angles the dependence of the differential 
cross section on the beam momentum was fitted and interpolated to the beam momentum of 
1.2~GeV/c. The resulting distribution was used as an input for the simulation of 
$dd \to \mathrm{^3He}n$. Figure~\ref{fig:norm}b shows the match of the angular 
distribution of $\mathrm{^3He}$ in data and the Monte-Carlo filtered event generator. 
The extracted integrated luminosity is determined to be
$L^1_{int}\:=\:(877 \pm 2_{stat} \pm 62_{sys} \pm 62_{norm})\,\mathrm{nb^{-1}}$, 
where the superscript 1 refers to the first part of the run. The systematic 
uncertainty reflects different parametrizations of the reference cross sections. In 
addition, the uncertainty of 7$\%$ in the absolute normalization of the reference data
is also included.
The result for the total cross section given below is based only on this
first part of the run. The second part was optimized for high luminosities and also served 
as a pilot run for a measurement of $dd \to \mathrm{^4He}\pi^{0}$. It provided data to 
extract high statistics differential distributions for $dd \to \mathrm{^3He}n\pi^{0}$. These 
have been absolutely normalized relative to the first part of the run using the rates of 
the $dd \to \mathrm{^3He}n\pi^{0}$ reaction. The integrated luminosity obtained for the 
second part of the run amounts to 
$L^2_{int}\:=\:(4909 \pm 13_{stat} \pm 350_{sys} \pm 350_{norm})\,\mathrm{nb^{-1}}$.

The uncertainty on the integrated luminosity (in total 10$\%$ if all contributions
are added quadratically) is the dominant source for the systematic error on the 
absolute normalization. Another source is associated with the cut on the 
cumulated probability distribution of the kinematic fit. In order to quantify
the influence of this cut, the analysis was repeated for different regions in the 
probability distribution. For the total cross section the maximum deviation from
the average value was taken as error. Changes in the shape of the differential 
distributions were extracted similarly, however excluding the variation in
the absolute scale. For all other analysis conditions according to the 
criteria discussed in Ref.~\cite{Barlow10} no significant systematic effect
was observed.

\section{Phenomenological models}
Presently, no theoretical calculation exists for a microscopic description 
of the investigated reaction. However, in order to have a sufficiently precise 
acceptance correction a model which reproduces the experimental data 
reasonably well is required. The ansatz used here is the incoherent
sum of a quasi-free reaction mechanism based on $dp\to\mathrm{^3He}\pi^0$ and a 
partial-wave expansion for the 3-body reaction. While the latter is limited 
to  $s$- and $p$-waves, the large relative momenta between the spectator 
nucleon and the rest system in the quasi-free model corresponding to 
higher partial waves motivate the incoherent sum and the neglection of
interference terms. 

\begin{figure*}
  \begin{center}
  \includegraphics[width=\columnwidth]{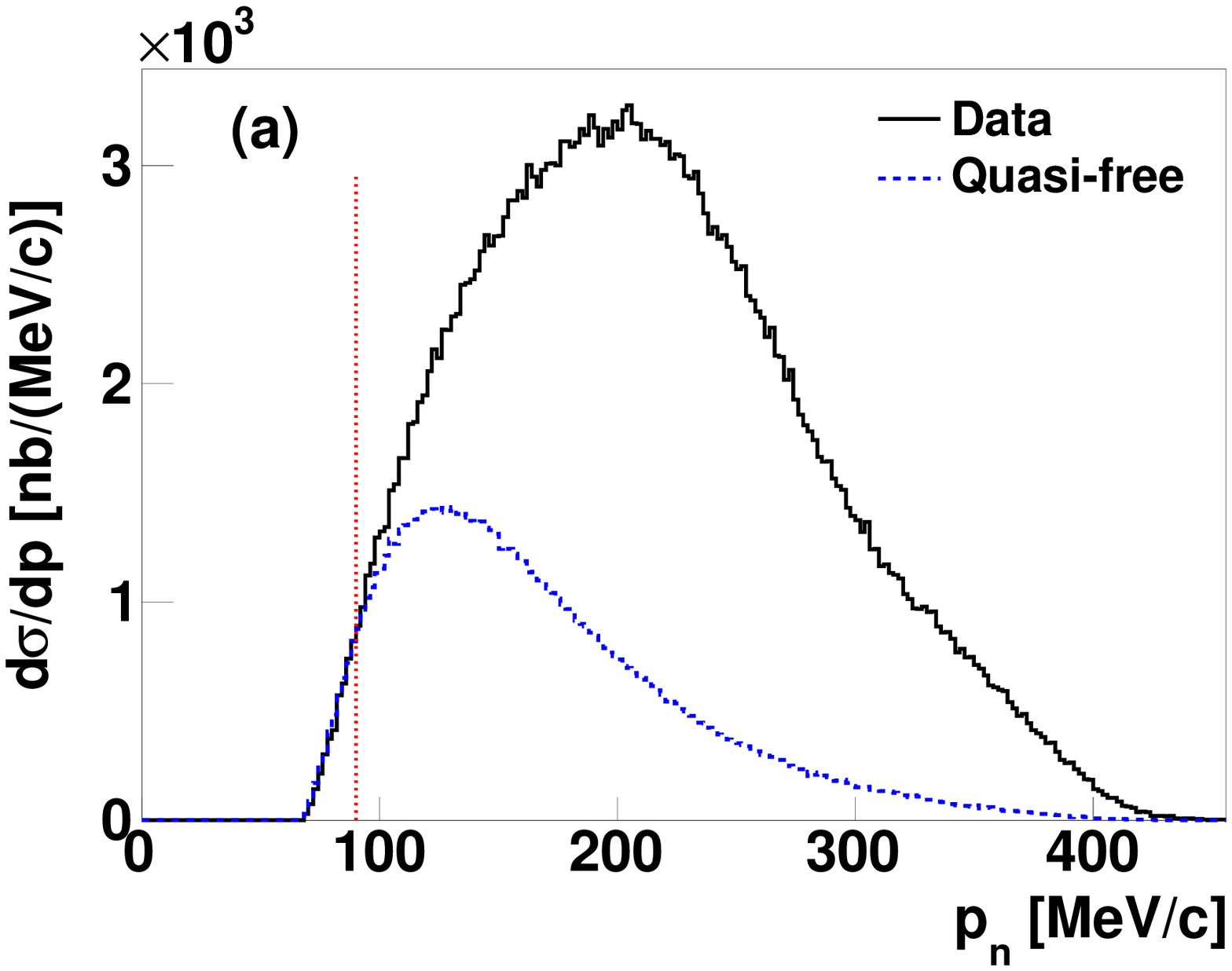} 
  \hfill
  \includegraphics[width=\columnwidth]{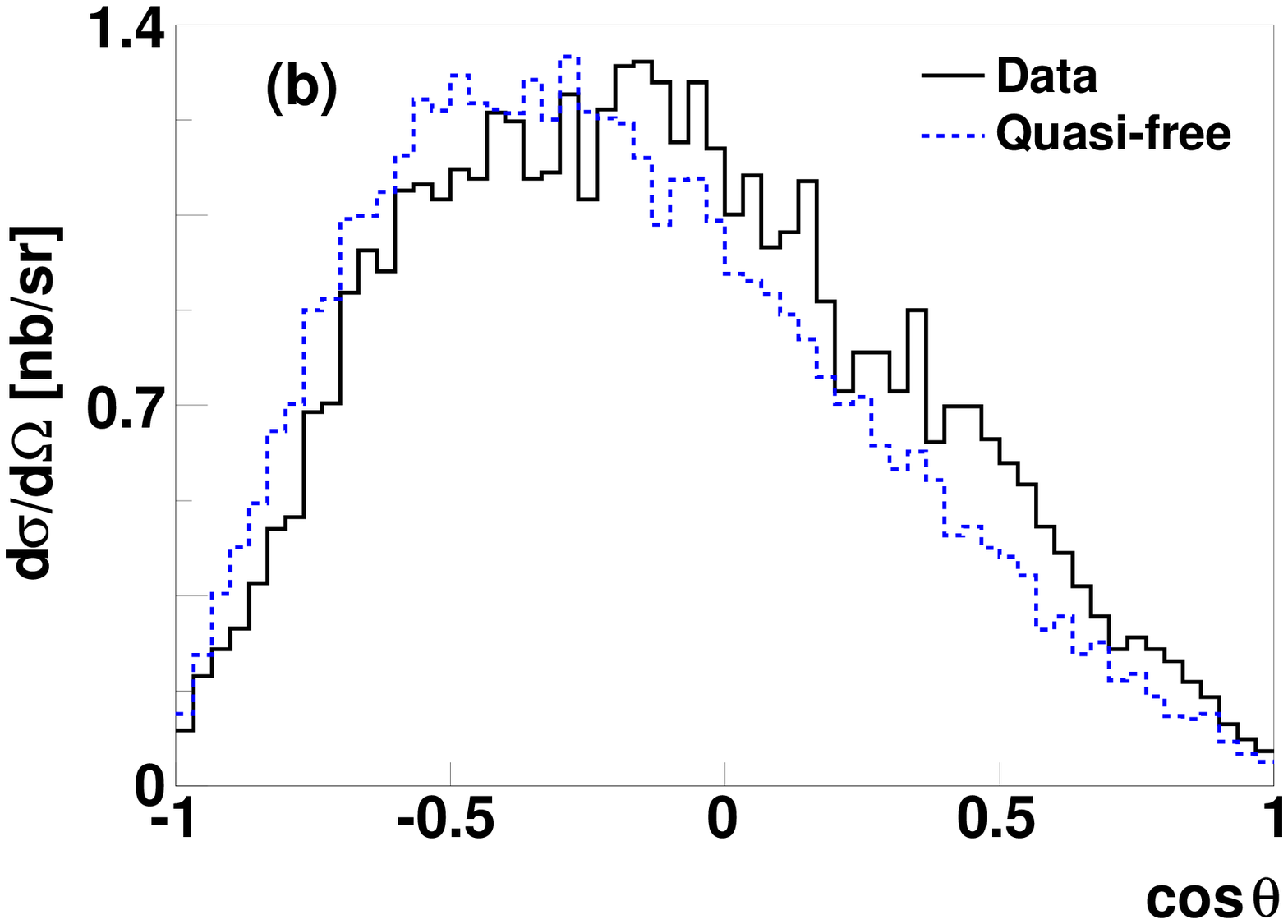}
  \end{center}
\caption{(Color online) Comparison of data (black line) and the quasi-free model
filtered by Monte-Carlo (blue dotted curve). (a) Momentum distribution of the 
neutron. (b) Angular distribution of the pion in the $\mathrm{^3He}-\pi^0$
subsystem for neutron momenta smaller than 90 MeV/c (indicated by the vertical 
red dotted line in the upper plot). Data are not corrected for acceptance.}
\label{fig:quasi-free}
\end{figure*}

\subsection{Quasi-free reaction model}

High momentum transfer reactions involving a deu\-teron can proceed via the 
interaction with a single nucleon of the deuteron and with the second 
nucleon being regarded as a spectator. Naturally, this mechanism is 
most significant in regions of the phase space where the momentum of 
one nucleon in final state matches the typical Fermi momenta in
the deuteron. In the present experiment two deuterons are 
involved and, thus, the reaction may proceed with a 
projectile or target neutron spectator. 
For the parametrization of the quasi-free sub reaction $dp\to\mathrm{^3He}\pi^0$, the 
empirical angular distributions and the energy dependent cross section 
in the energy regime from threshold up to an excess energy of 10 MeV \cite{Nikulin96} 
and for excess energies of 40, 60 and 80 MeV \cite{Betigeri01} have been used.
They have been convoluted with the momentum distribution of the proton 
in the deuteron using an analytical form of the deuteron wave function based 
on the Paris potential \cite{Lacombe81}. As a result one gets absolutely normalized 
differential cross sections for the quasi-free contribution to $dd\to\mathrm{^3He}n\pi^0$,
which can be directly compared to the measured data. Figure~\ref{fig:quasi-free}a
shows the momentum distribution of the neutron for data and the quasi-free
model filtered by Monte-Carlo. As expected the quasi-free process dominates
the distribution for small momenta.
The lower boundary of the spectrum
is caused by kinematic effects. At a beam momentum of 1.2~GeV/c
the reaction $dp\to\mathrm{^3He}\pi^0$ with the target proton at rest is below threshold and can only
occur for $p_\mathrm{fermi}>48$~MeV/c. The vanishing acceptance
at $\theta_\mathrm{^3He}<3^\circ$ further increases the
minimum Fermi momentum.
Figure~\ref{fig:quasi-free}b shows the angular distribution of the pion 
in the $\mathrm{^3He}-\pi^0$ subsystem for neutron momenta below 90 MeV/c, {\it i.e.} in
the region where the quasi-free process should dominate the distribution. 

\subsection{Partial wave decomposition}
For the remaining part of the data which cannot be described with the quasi-free process
a 3-body model based on a partial wave decomposition has been developed. 
The relative angular momenta were defined according to the coordinates introduced 
earlier: one in the global $\pi^{0}$ - ($\mathrm{^3He}n$) system and one in the 
$\mathrm{^3He}-n$ subsystem (denoted by $l$ and $L$, respectively). For the partial wave 
decomposition the angular momenta have been limited to $l + L \leq 1$, {\it i.e.} 
to at most one $p$-wave in the system. For the momentum dependence the standard 
approximation $|M|^2 \propto q^{2l} p^{2L}$ was used. Taking into account all possible 
spin configurations this results in 18 possible amplitudes. After combining the amplitudes 
with the same signature in final state, four possible contributions can be identified: 
$s$-wave in both systems ($sS$), one $p$-wave in either system ($sP$ and $pS$) and a 
$sP-pS$ interference term. They can be described by seven real coefficients 
(four complex amplitudes minus one overall phase). 
With this the four-fold differential cross section can be written as:
\begin{widetext}
\begin{equation} 
\begin{array}{rcl}
\displaystyle \frac{\mathrm{d}^4\sigma}{2\pi~\mathrm{d}M_{^{3}\mathrm{He}n}~\mathrm{d}\cos\theta_p~\mathrm{d}\cos\theta_q~\mathrm{d}\varphi} &=
C\,pq & \left[ \frac{}{} A_0 + A_1 q^2 +  A_3 p^2 + \frac{1}{4}A_2 q^2 \left(1+3\cos 2\theta_q \right) + 
\frac{1}{4}A_4 p^2 \left(1+3\cos 2\theta_p \right) \right. \\
\displaystyle && \left. +  A_5 p q \cos\theta_p \cos\theta_q + A_6 p q \sin\theta_p \sin\theta_q\cos\varphi \frac{}{} \right]\:
\end{array}
\label{d4sigma}
\end{equation}
\end{widetext}
with
\begin{equation}
C = \frac{1}{32(2\pi)^5 sp_a^* (2s_a+1)(2s_b+1)}
\end{equation}
where $s_a$ and $s_b$ denote the spin of beam and target and $s$ and $p_a^*$ the total energy squared 
and the beam momentum, respectively, in the c.m. system. The coefficients $A_i$ describe the 
strength of the individual contributions mentioned above: $A_0$ corresponds to $l=L=0$ ($sS$), 
$A_1$ and $A_2$ to $l=1$ and $L=0$ ($pS$), $A_3$ and $A_4$ to $l=0$ and $L=1$ ($sP$)
and $A_5$ and $A_6$ to the interference term.
Integration of Eq.~\ref{d4sigma} results in a set of equations for the description 
of the single differential cross sections:
\begin{subequations}
\label{eq:whole}
\begin{eqnarray} 
\frac{\mathrm{d}\sigma}{\mathrm{d}M_{^{3}\mathrm{He}n}} & = & 16\pi^2 Cpq\left[A_0+A_1q^2+A_3p^2\right]\:
\\
\frac{\mathrm{d}\sigma}{2\pi \mathrm{d}\cos\theta_q} & = &4\pi
C\left[B+\frac{1}{4}A_2\left(1+3\cos 2\theta_q
\right)I_{pS}\right]\:
\\
\frac{\mathrm{d}\sigma}{2\pi \mathrm{d}\cos\theta_p} & = & 4\pi
C\left[B+\frac{1}{4}A_4\left(1+3\cos 2\theta_p
\right)I_{sP}\right]\:
\\ 
\frac{\mathrm{d}\sigma}{\mathrm{d}\varphi}& = &8\pi
C\left[B+\frac{\pi^2}{16}A_6I_{pS+sP}\cos\varphi\right]\:
\end{eqnarray}
\end{subequations}
with the new coefficient 
\begin{equation}
\label{eq:bcoeff}
B=A_{0}I_{sS}+A_{1}I_{pS}+A_{3}I_{sP}.
\end{equation}
The constants $I_{sS}$, $I_{pS}$, $I_{sP}$ and $I_{pS+sP}$ are the results of 
the integration over $M_{\mathrm{^3He}n}$:
\begin{subequations}
\label{eq:intcontst}
\begin{eqnarray} 
I_{sS}&=&\int_{(M_{\mathrm{^{3}He}}+M_n)^2}^{(\sqrt{s}-M_\pi)^2}pq\,\mathrm{d}M_{^{3}\mathrm{He}n}\:
\\
I_{pS}&=&\int_{(M_{\mathrm{^{3}He}}+M_n)^2}^{(\sqrt{s}-M_\pi)^2}pq^3\,\mathrm{d}M_{^{3}\mathrm{He}n}\:
\\
I_{sP}&=&\int_{(M_{\mathrm{^{3}He}}+M_n)^2}^{(\sqrt{s}-M_\pi)^2}p^3q\,\mathrm{d}M_{^{3}\mathrm{He}n}\:
\\
I_{pS+sP}&=&\int_{(M_{\mathrm{^{3}He}}+M_n)^2}^{(\sqrt{s}-M_\pi)^2}p^2q^2\,\mathrm{d}M_{^{3}\mathrm{He}n}\:
\end{eqnarray}
\end{subequations}

Equations~\ref{eq:whole} do not contain the coefficient $A_5$ as the corresponding term
vanishes with the integration over $\cos\theta_{q}$ and $\cos\theta_{p}$. In order to extract 
this coefficient Eq.~\ref{d4sigma} has to be multiplied by $\cos\theta_q\cos\theta_p$ before
integration. This results in the following formula to determine $A_5$:
\begin{equation}
\label{eq:phimod}
\frac{\mathrm{d}\sigma'}{\mathrm{d}\varphi}=\frac{8}{9}\pi C A_5 I_{pS+sP}
\end{equation}
with $\sigma'(q,p) = \sigma(q,p) \cdot \cos\theta_q\cos\theta_p$.

It has to be noted that the coefficients $A_{0}$, $A_{1}$ and $A_{3}$ cannot be extracted 
unambiguously from the differential distribution $\mathrm{d}\sigma/\mathrm{d}M_{\mathrm{^3He}n}$. In the
non-relativistic limit $q^2$ and $p^2$ are both linear in $M_{\mathrm{^3He}n}$ introducing
a correlation of all three coefficients. For the measurement of $dd \to \mathrm{^3He}n\pi^{0}$ 
at an excess energy of $Q\approx\,$40 MeV a non-relativistic treatment is still a good approximation. Thus, only a
value for B can be extracted from the data. Any values for $A_{0}$, $A_{1}$ and $A_{3}$
fulfilling Eq.~\ref{eq:bcoeff} and the fit to $\mathrm{d}\sigma/\mathrm{d}M_{\mathrm{^3He}n}$ will lead 
to the same model description. However, in order to provide a complete set of
coefficients the parameter $A_1$ has been fixed manually.

\section{Results}

\begin{figure*}
\begin{center}
\includegraphics[width=\columnwidth]{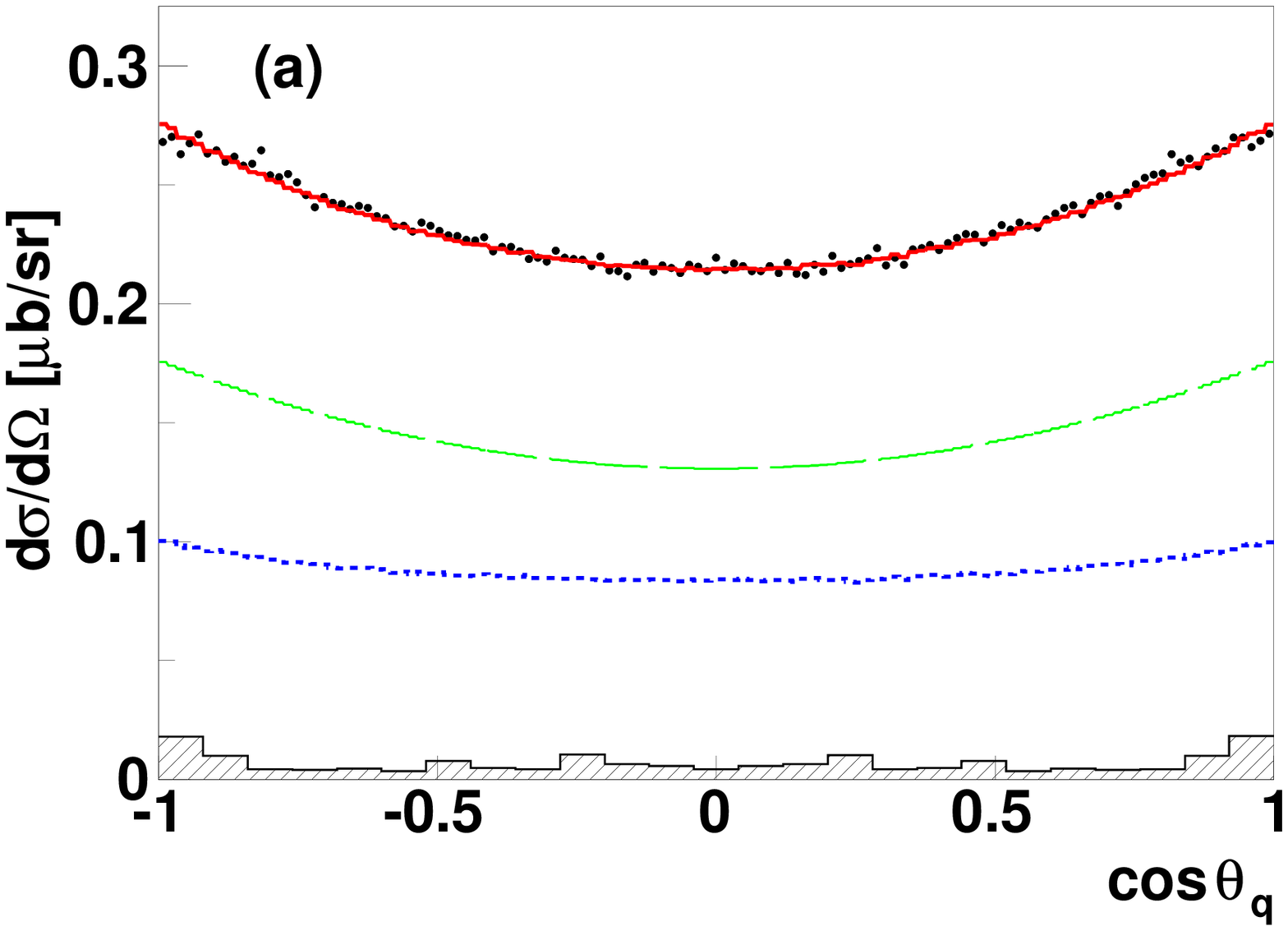}
\hfill 
\includegraphics[width=\columnwidth]{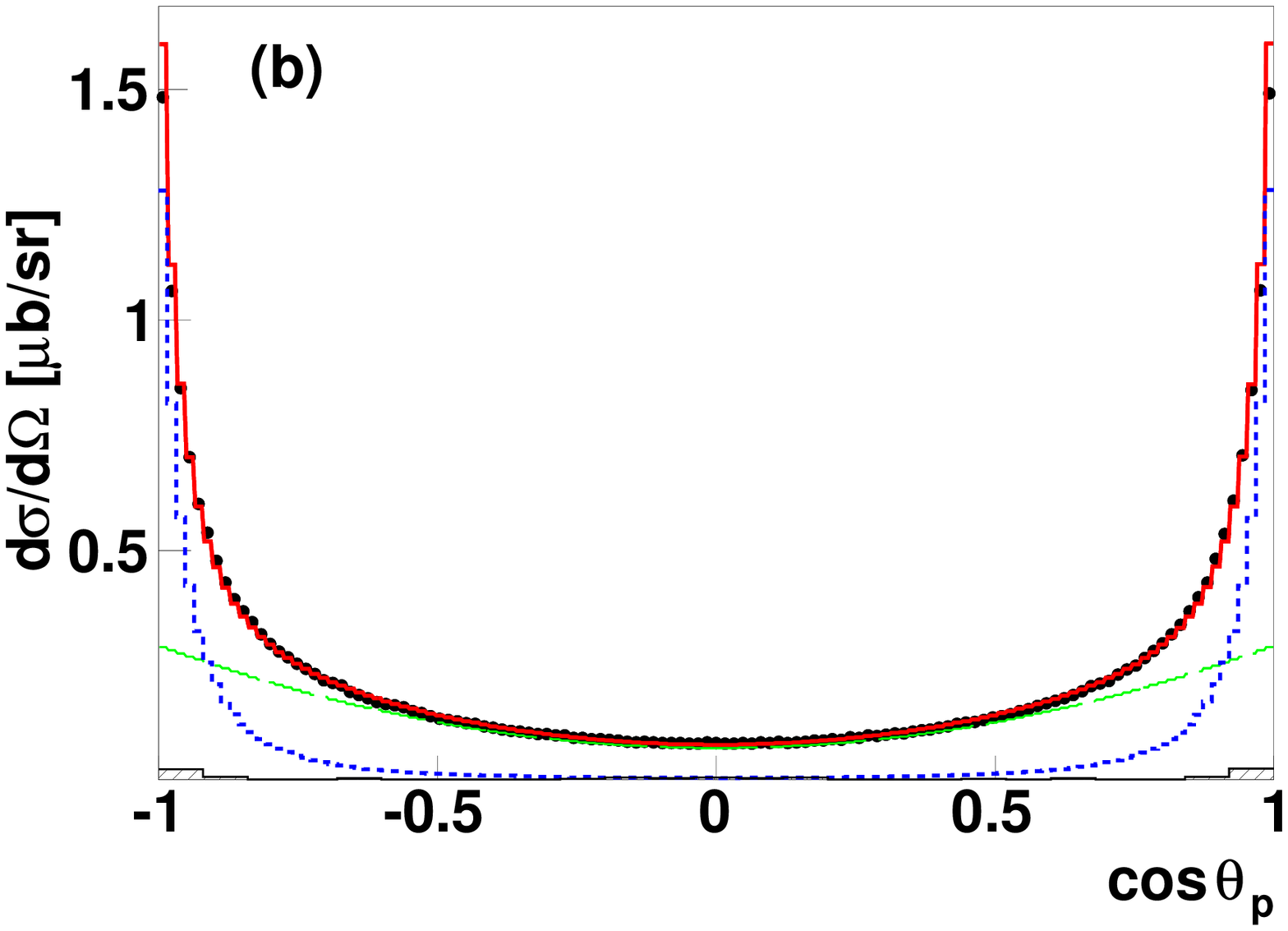} 
\includegraphics[width=\columnwidth]{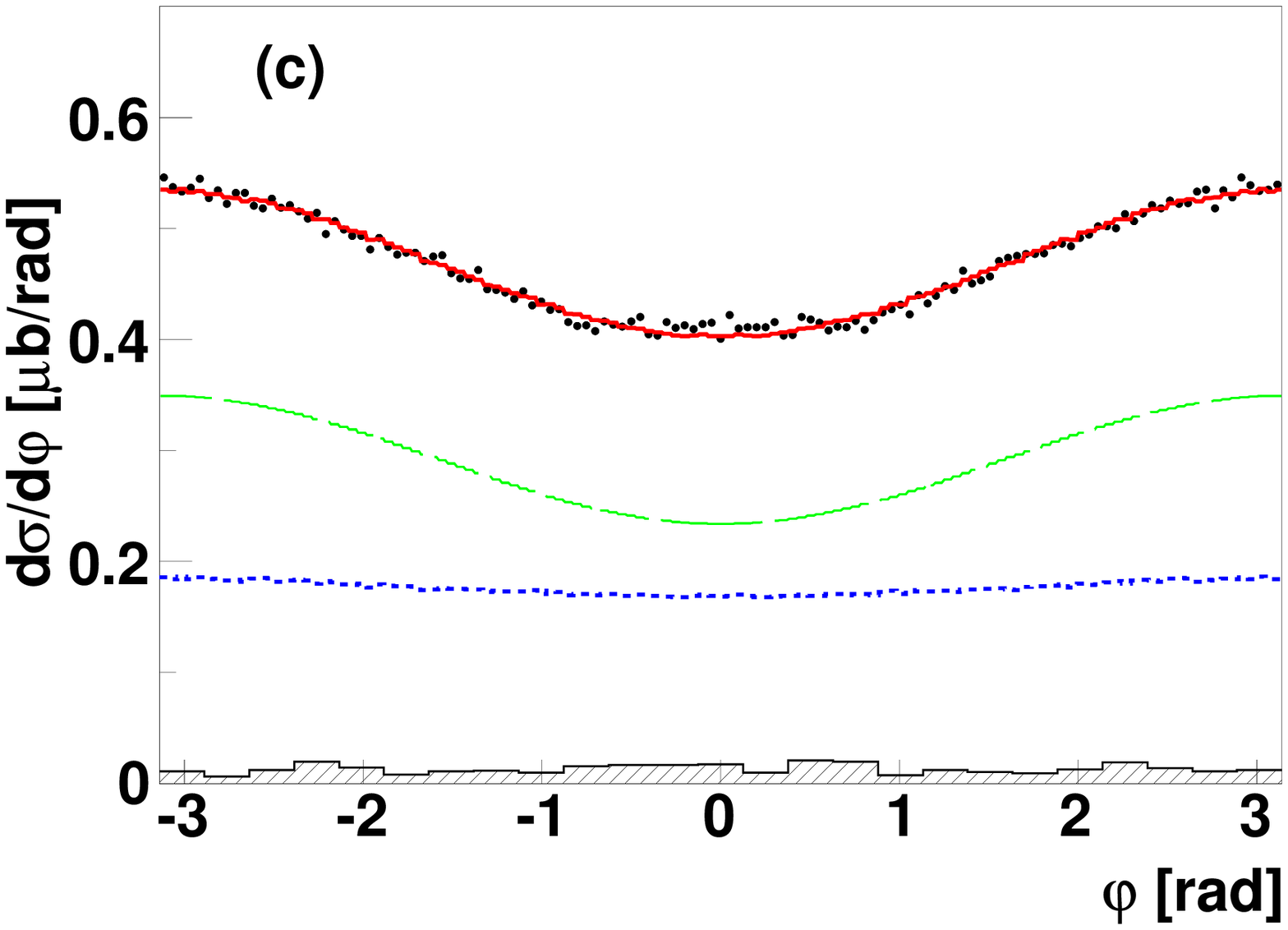} 
\hfill
\includegraphics[width=\columnwidth]{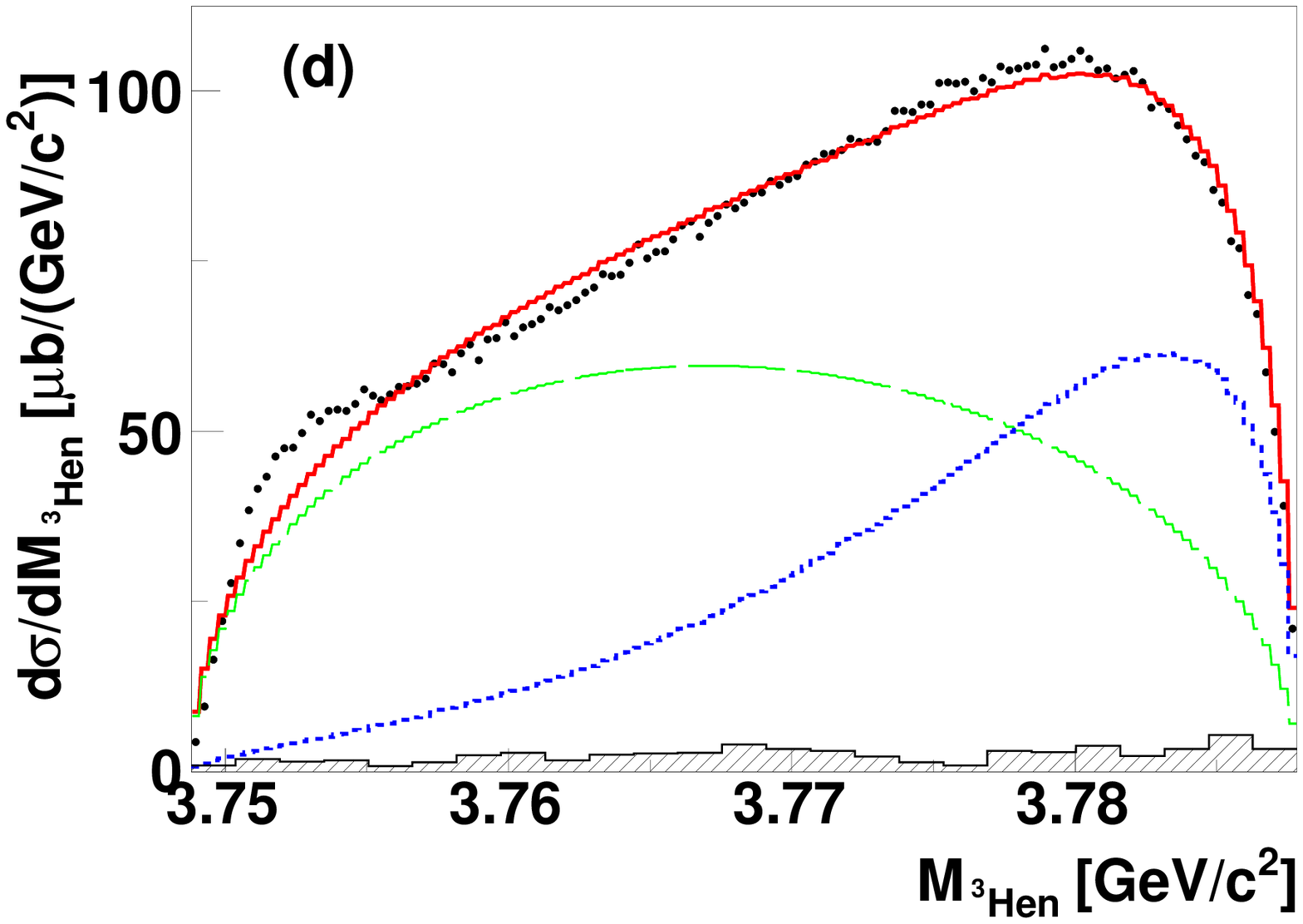} 
\end{center}
\caption{(Color online) Acceptance corrected data (black points) presented as functions 
of (a) $\cos\theta_{q}$, (b) $\cos\theta_{p}$, (c) $\varphi$ and (d) $M_{^{3}\mathrm{He}n}$. 
The curves represent the fit to the model: full model (red solid), quasi-free 
contribution (blue dotted) and the partial wave decomposition (green long dashed). 
The hatched areas indicate the systematic uncertainties on the shape of the differential 
distributions. Uncertainties on the absolute normalization are not included.
}
\label{fig:diff_4}       
\end{figure*}

 In a first step a sum of Monte-Carlo filtered distributions for each contribution from 
the partial wave decomposition (coefficients $A_0$ to $A_6$) and from the quasi-free model 
(coefficient $A_7$) was fitted to the uncorrected, single differential spectra. The 
result served as input for the Monte-Carlo simulation finally used to determine
the acceptance correction. 

\begin{figure}
\begin{center}
\includegraphics[width=\columnwidth]{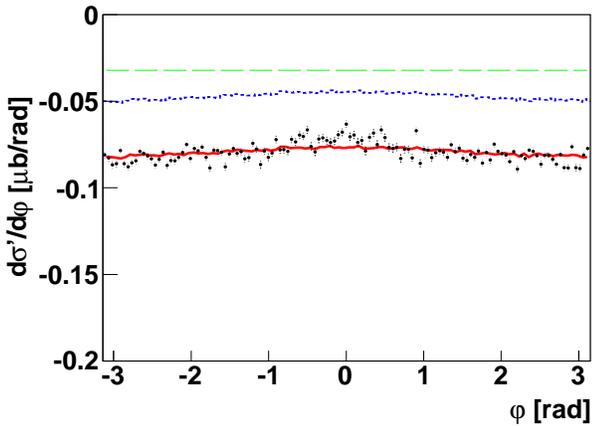}
\end{center}
\caption{(Color online) Distribution of $\mathrm{d}\sigma'/\mathrm{d}\varphi$ as used in Eq.~\ref{eq:phimod} to extract $A_5$.
For the definition of the curves see Fig.~\ref{fig:diff_4}.}
\label{fig:diff_5}       
\end{figure}

The final distributions after acceptance correction are presented in Fig.~\ref{fig:diff_4}. 
Contributions from the quasi-free model, the partial wave decomposition and the full 
model are shown in blue, green and red, respectively. These spectra were refitted 
using the analytical formulas given in the previous section. The result is consistent
with the initial fit. Although the partial wave expansion was limited to at most 
one $p$-wave in the final state it provides a reasonable overall description of the 
data: both angular distributions show a significant contribution of $p$-waves of 
similar size, the $pS$-$sP$ interference term is visualized by the non-isotropic
distribution of $\mathrm{d}\sigma/\mathrm{d}\varphi$. The quasi-free contribution is about $1.11\,\mathrm{\mu b}$ 
and, thus, is in agreement with the prediction of the quasi-free model ($1.19\,\mathrm{\mu b}$) 
within the normalization error of about $10\%$ given in Ref.~\cite{Nikulin96}. 
The result of the fit using Eq.~\ref{eq:phimod} and the quasi-free model is 
presented in Fig.~\ref{fig:diff_5}. 
The values for the extracted coefficients from the global fit are summarized in Table~\ref{coeftab}.

\begin{table}[b]
\begin{center}
\begin{tabular}{lrd{3.3}cd{1.3}cll}
\hline
\hline
Parameter & \multicolumn{7}{c}{Fit result} \\
\hline
$B$     & ( & 1.840 & $\pm$ & 0.003 &)&           &$\mathrm{\mu b}$\\ 
$A_{0}$ & ( & 0.41  & $\pm$ & 0.01  &)&$\cdot10^4$&$\mathrm{\mu b/GeV^{3}}$\\ 
$A_{1}$ &   & 8.4   &       &       & &$\cdot10^4$ &$\mathrm{\mu b/GeV^{5}}$\\
$A_{2}$ & ( & 18.3  & $\pm$ & 0.3   &)&$\cdot10^4$&$\mathrm{\mu b/GeV^{5}}$\\ 
$A_{3}$ & ( & 1.08  & $\pm$ & 0.05  &)&$\cdot10^4$&$\mathrm{\mu b/GeV^{5}}$\\
$A_{4}$ & ( & 18.04 & $\pm$ & 0.07  &)&$\cdot10^4$&$\mathrm{\mu b/GeV^{5}}$\\ 
$A_{5}$ & ( & -45.4 & $\pm$ & 0.3   &)&$\cdot10^4$&$\mathrm{\mu b/GeV^{5}}$\\
$A_{6}$ & ( & -15.0 & $\pm$ & 0.2   &)&$\cdot10^4$&$\mathrm{\mu b/GeV^{5}}$\\
$\sigma_{qf}\cdot A_{7}$ & ( & 1.108 & $\pm$ &  0.003  & ) &&$\mathrm{\mu b}$\\
\hline
\hline
\end{tabular}
\caption{Collection of the extracted fit parameters. The amplitudes are given in units of 
$(4\pi)^{2}C$. The parameters $A_0$, $A_1$ and $A_3$ are correlated and could not be
extracted unambiguously: the given numbers represent one
possible solution with $A_1$ being fixed (see text).}
\label{coeftab}
\end{center}
\end{table}

One should emphasize that the meaning of the fit parameters is limited to the context
of the discussed model. Any addition of higher partial waves, for example, would
also change the extracted amplitudes of the lower partial waves. Thus, systematic errors
are only provided for data. For the extracted fit parameters only statistical
errors are given.

\begin{figure}
\begin{center}
\includegraphics[width=\columnwidth]{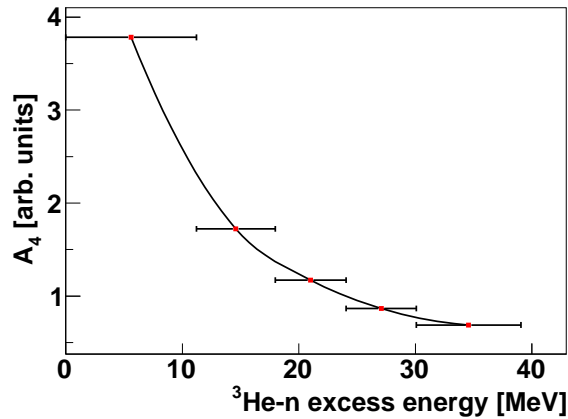}
\end{center}
\caption{Coefficient $A_4$ in the 
$^{3}\mathrm{He}-n$ system as a
function of the excess energy: the strength of the $p$-wave contribution
increases for small excess energies. The error bars along the $x$-axis represent
the width of the intervals in $M_{^{3}\mathrm{He}n}$.
}
\label{fig:qdep}       
\end{figure}

So far, a possible momentum dependence of 
the transition amplitudes, for example due to initial or final state interaction, was neglected.
Deviations from this assumption were studied by refitting the data for five
intervals in $M_{^{3}\mathrm{He}n}$ (corresponding to intervals in $\vec{q}$ and $\vec{p}$).
All coefficients except one remained constant. Only $A_4$ representing a $p$-wave contribution 
in the $^{3}\mathrm{He}n$ system showed a significant momentum dependence (see Fig.~\ref{fig:qdep}):
$A_4$ is larger for low excess energies in the $^{3}\mathrm{He}n$ system (corresponding to low 
relative momenta). One possible
reason for this might be excited states with isospin $I=1$ in the $^{3}\mathrm{He}n$ system 
at low excess energies as reported in Ref.~\cite{Tilley92} (the production of an $I=0$ state
would be charge symmetry breaking). 

\begin{figure}
\begin{center}
\includegraphics[width=\columnwidth]{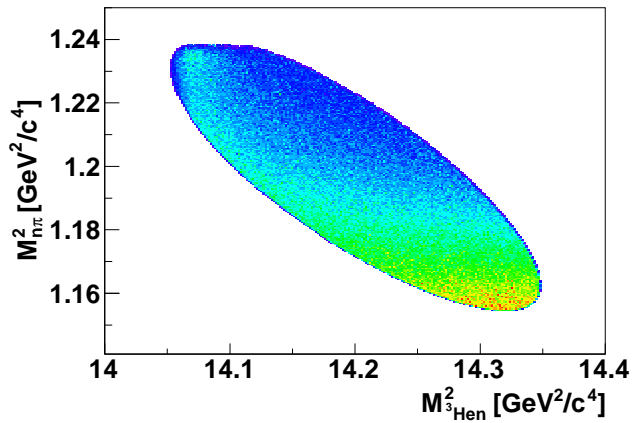}
\end{center}
\caption{(Color online) Acceptance corrected Dalitz plot. The region with low values 
of $M^2_{n\pi}$ and large values of $M^2_{\mathrm{^3He}n}$ is dominated by the 
quasi-free process. The enhancement at low values of  $M^2_{\mathrm{^3He}n}$ corresponds 
to the low mass enhancement in Fig.~\ref{fig:diff_4}d and might be connected to
the energy dependence of the $p$-wave amplitude discussed in Fig~\ref{fig:qdep}. At large values of 
 $M^2_{n\pi}$ the Dalitz plot is cut due to the acceptance hole for $\theta_{^{3}\mathrm{He}}<3^\circ$. 
}
\label{fig:dalitz}
\end{figure}

Figure~\ref{fig:dalitz} shows the acceptance corrected Dalitz plot for $M^2_{n\pi}$ 
versus $M^2_{\mathrm{^3He}n}$. It should be noted 
that the Dalitz plot is fully covered except for a small region for large $\pi^{0}-n$ 
invariant masses due to the acceptance hole for $\theta_{^{3}\mathrm{He}}<3^\circ$.
The quasi-free reaction process mainly populates the region for small
 $\pi^{0}-n$ invariant masses and large $\mathrm{^3He}-n$ invariant masses. 
The observation of an increasing $p$-wave contribution for small excess
energies in the $\mathrm{^3He}-n$ system possibly caused by an excited 
$I=0$ state comes with an enhancement in the Dalitz plot for small 
$\mathrm{^3He}-n$ invariant masses.

Integrating over the differential distributions we obtain for the total cross 
section of the $dd \rightarrow \mathrm{^3He}n\pi^0$ reaction: 
\begin{eqnarray}
\label{eq:mysigtot} 
\sigma_{tot}= (2.89 \pm 0.01_{stat}\pm 0.06_{sys} \pm 0.29_{norm})\,\mathrm{\mu b}.
\end{eqnarray}

\section{Summary}
For the first time an exclusive measurement of the $dd \rightarrow \mathrm{^3He}n\pi^0$ reaction 
has been performed. A total cross section of $\sigma_{tot}= 2.89\,\mathrm{\mu b}$ with an 
accuracy of about 11$\%$ has been extracted. Differential distributions have
been compared to the incoherent sum of a quasi-free reaction model and a 
partial-wave expansion limited to at most one $p$-wave in the final state.
The contribution of the quasi-free processes accounts for about $1.11\,\mathrm{\mu b}$ of 
the total cross section matching the prediction of the quasi-free reaction model.
The partial wave decomposition reveals the importance of $p$-wave contributions 
in the final state. The applied model shows a reasonable agreement for all 
differential distribution. Thus, based on this comparison no indication
for significant contributions of higher partial waves can be deduced.

The whole data set amounts to about $3.4\times10^{6}$ fully reconstructed and 
background-free events. The presented differential distributions are only one possible
representation of the results. One goal of the measurement was to provide
data for studying $dd$ initial state interaction for small angular momenta,
which is one missing information in the microscopic description of
the charge symmetry breaking reaction $dd \to \mathrm{^4He}\pi^0$ within the 
framework of Chiral Perturbation Theory. Once the important observables
have been identified the corresponding experimental distributions can be
provided.
\section{Acknowledgments}
We would like to thank the technical and administrative staff
at the Forschungszentrum J\"{u}lich, especially at the COoler SYnchrotron
COSY and at the participating institutes. This work has been supported in part by the 
German Federal Ministry of
Education and Research (BMBF), the Polish Ministry of Science and Higher
Education, the Polish National Science Center (grant No. 2011/01/B/ST2/00431), the Foundation for Polish Science (MPD),
Forschungszentrum J\"{u}lich (COSY-FFE) and the European Union
Seventh Framework Programme (FP7/2007-2013) under grant agreement No. 283286.

\end{document}